# Direct Bandgap Photoluminescence of GeSn grown on Si(100) substrate by Molecular Beam Epitaxy Growth

Diandian Zhang,[1] Nirosh M. Eldose,[2] Dinesh Baral,[1,2] Hryhorii Stanchu,[2] Sudip Acharya,[1] Fernando Maia de Oliveira,[2] Mourad Benamara,[1] Haochen Zhao,[3] Yuping Zeng,[3] Wei Du,[1,2] Gregory J. Salamo,[2,4] and Shui-Qing Yu[1,2*]

[1] *Department of Electrical Engineering and Computer Science, University of Arkansas, Fayetteville, Arkansas, 72701, USA*

[2] *Institute for Nanoscience and Engineering, University of Arkansas, Fayetteville, Arkansas, 72701, USA*

[3] *Department of Electrical and Computer Engineering, University of Delaware, Newark, Delaware 19716, USA*

[4] *Department of Physics, University of Arkansas, Fayetteville, Arkansas, 72701, USA*

*Corresponding Author: syu@uark.edu

## Abstract

Group IV alloys of GeSn have gained significant attention for electronic and optoelectronic applications on a Si platform due to their compatibility with existing CMOS technology, tunable band structure, and potential for a direct bandgap at high Sn concentrations. However, synthesizing Sn-rich $Ge_{1-x}Sn_x$ structures remains challenging due to the low solid solubility of Sn in Ge (< 1%) and the substantial lattice mismatch (~14%) between α-Sn and Ge. In this work, we demonstrate the successful growth of high-quality, relaxed GeSn layers with Sn contents of 9.2% and 11.4% on Si(100) substrates via molecular beam epitaxy (MBE). As far as we know, this is the first report of direct bandgap photoluminescence observed from MBE-grown GeSn films without post-growth annealing. Structural characterizations including X-ray diffraction (XRD), secondary ion mass spectrometry (SIMS), and transmission electron microscopy (TEM) confirm uniform Sn incorporation with minimal defect formation. Atomic force microscopy (AFM) reveals smooth surfaces with low roughness. Temperature-dependent photoluminescence (PL) measurements further confirm direct bandgap emission, representing a new stage in the development of MBE-grown GeSn.

## Introduction

Germanium–tin (GeSn) alloys have emerged as highly promising candidates for silicon-based photonics and optoelectronics due to their compatibility with existing CMOS technology and their composition-tunable band structure. Incorporating approximately 6–10% Sn into Ge enables a transition from an indirect to a direct bandgap, which promise efficient radiative recombination[1-4]. This unique property and the prospect of monolithic integration on Si make GeSn attractive for applications in shortwave infrared (SWIR) and mid-infrared (MIR) light emission, photodetection, and silicon-integrated lasers[5-7]. Moreover, the compatibility of GeSn with metal–oxide–semiconductor (CMOS) technology further enhances its application potential. However, achieving high-quality GeSn films with sufficient Sn incorporation while maintaining high crystal quality remains a significant challenge, especially for MBE, particularly due to strain driven defect formation and Sn segregation during growth[8-10].

MBE is a powerful technique for the growth of GeSn films due to its precise control of composition and interface quality at low growth temperatures[11, 12]. Moreover, MBE allows for uniform Sn incorporation and strain engineering, potentially facilitating the realization of high-Sn-content GeSn layers with enhanced optical properties[13, 14]. Despite the potential of MBE, strain relaxation in GeSn films grown on Si(100) substrates remains a critical issue, as the large lattice mismatch between GeSn and Si leads to strain-relaxation induced defects and Sn segregation that degrade PL efficiency[15-19]. As a result, while previous studies have demonstrated high PL emission from GeSn films grown by chemical vapor deposition (CVD), to date, there has been no reports of direct bandgap PL from GeSn grown by MBE[6, 7, 20-23]. In this paper, we report the first observation to our knowledge of direct bandgap photoluminescence from GeSn films grown on a Si(100) substrate using MBE, without post-annealing, marking a significant contribution to the development of high-Sn-content GeSn. Comprehensive structural and optical characterizations, XRD, SIMS, TEM, AFM, and temperature-dependent PL, confirm successful Sn incorporation up to ~11.4%, effective strain relaxation, and strong direct bandgap photoluminescence from GeSn films. These results mark a crucial step forward in the development of high-Sn-content GeSn grown by MBE.

**Experiment**

GeSn layered structure was grown on a 2-inch wafer Si(100) utilizing the MBE with Knudsen cells made of pyrolytic boron nitride (PBN) crucibles, loaded with ultra-high (7N) pure intrinsic Ge and metallic Sn respectively. The GeSn samples were grown in an ultra-high vacuum (UHV) MBE chamber at a background pressure of $1\times10^{-11}$ Torr. Prior to loading, Si substrates were chemically cleaned using a diluted HF:H$_2$O (1:20) solution, dipped for one minute, followed by nitrogen blow-drying. The substrates were subsequently degassed at 250 °C for over two hours inside the MBE chamber and then heated to ~900 °C to remove the oxide layer. A two-step Ge buffer layer was first deposited to improve surface morphology and accommodate lattice mismatch. The low-temperature Ge layer (~70 nm) was grown at 400 °C, followed by a high-temperature layer (~120 nm) at 700 °C. To enhance crystal quality, three cyclic annealing was performed at 700-800 °C. The substrate was then cooled to ~180 °C to grow the first GeSn layer, followed by the second GeSn layer grown at 200 and 220 ºC for sample S1 and S2, respectively. The schematic diagram of the GeSn samples is shown in Fig. 1(a).

High-resolution XRD measurements were performed on GeSn samples using a Panalytical X'Pert Pro MRD diffractometer equipped with a 1.8 kW Cu Kα1 X-ray tube (λ = 1.540598 Å), a standard four-bounce Ge(220) monochromator, and a Pixel detector. TEM analysis was performed using a FEI Titan 80–300 microscope operated at 300 KV. The PL data was acquired using a 1064 nm laser as an excitation source and a Horiba spectrometer coupled with a InSb photodetector. A 1200 nm long-pass filter was used to block light from the laser excitation beam.

**Results**

Fig. 1(b) shows the XRD $2\theta$-$\omega$ scans of GeSn samples, which reveal distinct diffraction peaks corresponding to the GeSn layer, Ge buffer layer, and Si substrate. The presence of a well-defined GeSn peak with a comparable FWHM to that of Ge buffer confirms the successful GeSn growth with a high quality. The large separation between the GeSn and Ge peaks on the $2\theta$-$\omega$ scan is due to the high Sn content. To further analyze the strain state, reciprocal space mapping (RSM) measurements were performed around the asymmetric ($\bar{2}\bar{2}4$) reflection, as shown in Fig. 1(c). The vertical and inclined dashed lines on the RSM represent fully strained and fully relaxed conditions. The magnitude of strain ($\varepsilon_{//}$) and the degree of strain relaxation (R) was estimated by measuring the in-plane ($a_{//} = (2\pi)2\sqrt{2}/Q_x$) and out-of-plane ($a_\perp = (2\pi)4/Q_z$) lattice parameters of the GeSn layer from the $Q_x$ and $Q_z$ coordinates of the GeSn peak on (224) RSM. Using Vegard's law and Poisson's relationship, the relaxation degrees were determined to be 47% and 68% for samples S1 and S2, respectively, and the Sn compositions were determined to be 9.2% and 11.4%, respectively. The sample with a lower Sn composition exhibited a lower degree of relaxation, which is attributed to the smaller thickness and lower Sn composition resulting in lower strain energy, leading to reduced relaxation, a phenomenon commonly observed in heteroepitaxy[24].

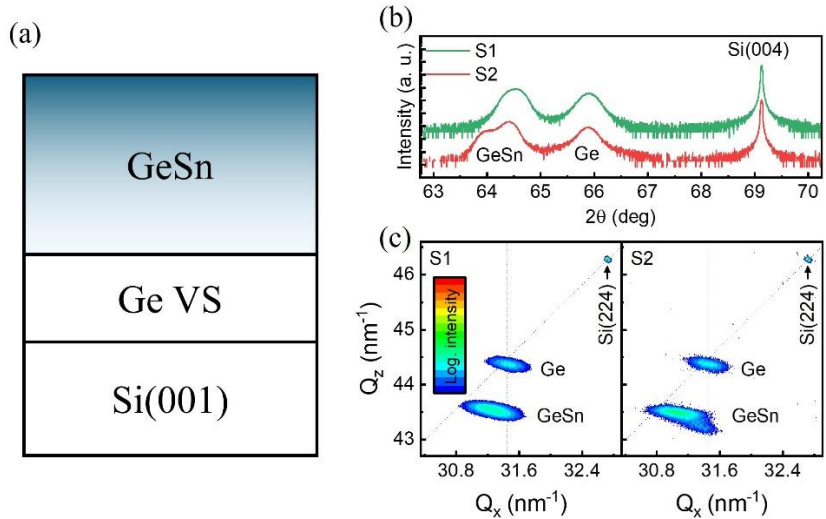

Figure 1. (a) Schematic diagram of the GeSn grown on Si(100) substrate. (b) XRD 004 2θ-ω scan of GeSn/Ge (VS)/Si(100) S1 (Sn: 9.2%) and S2 (Sn: 11.4%) samples, and (c) XRD-RSM of GeSn near Ge(224), showing strain relaxation.

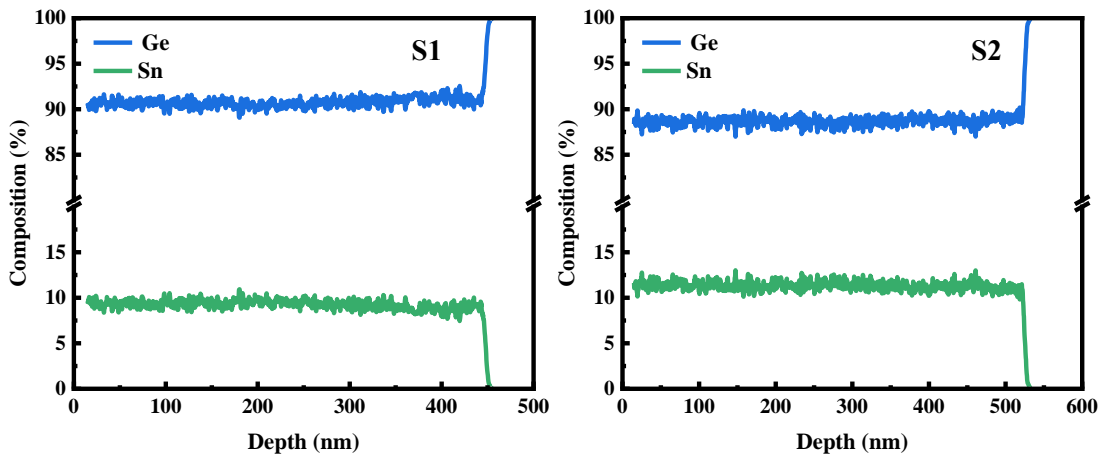

Figure 2. SIMS depth profile showing Ge and Sn concentration in the samples S1 and S2.

Fig. 2 shows the SIMS depth profile of sample S1 and S2, confirming a Sn content of 9.2% and 11.4% respectively, consistent with XRD-RSM measurements. Also, the SIMS profiles reveal the GeSn layer thickness of 450 and 525 nm for samples S1 and S2, respectively. The uniform Sn distribution indicates homogeneous incorporation and ensuring compositional consistency and structural integrity.

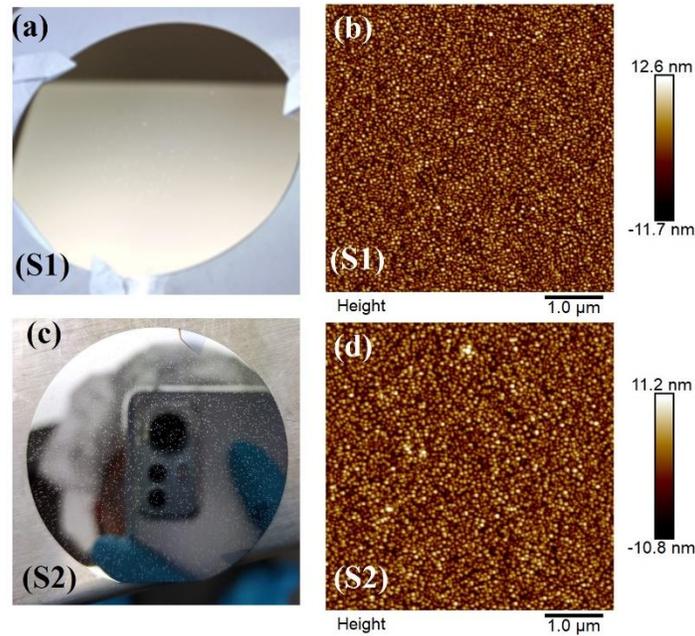

Figure 3. (a and c) Optical images of S1 (Sn:9.2%) and S2 (Sn:11.4%) respectively, and (b and d) AFM images of samples S1 and S2 respectively.

Fig. 3(a and c) show the optical images of samples S1 and S2, as shown in Fig. 3(a). Sample S1 exhibits a shiny surface, but a small number of white dot-like features can still be observed. These are traces left by Sn segregation, indicating a slight occurrence of Sn segregation. In contrast, sample S2, as shown in Fig. 3(c), also exhibits a shiny surface, but the segregation is more severe than in S1. This is attributed to the higher Sn content. However, it remains at a low segregation level, no high-density Sn droplets were observed, nor did the sample surface become rough or milky white. Overall, both samples exhibit a shiny surface, demonstrating the success of the epitaxy and the high quality of the epitaxial layers. In more detail, Fig. 3(b and d) shows the AFM analysis of samples S1 and S2. Both samples exhibit similar surface morphologies, with root mean square (RMS) surface roughness values of 3.08 nm and 2.62 nm, respectively, indicating a smooth morphology. The absence of Sn droplets on the images suggests efficient incorporation of Sn atoms into the GeSn lattice, with minimal segregation. These results confirm the high crystalline quality of the GeSn layer, which is essential for achieving optimal optical and electronic performance.

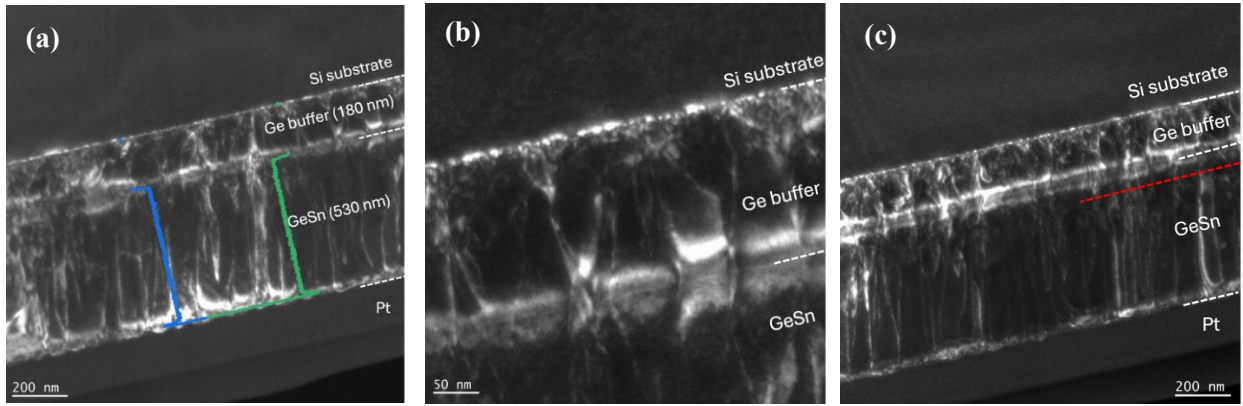

Figure 4. Dark field (DF) TEM images of sample S2 with Sn content of 11.4% at various magnifications. The blue and green lines in (a) are Ge and Sn concentration profile obtained by SIMS (Fig. 2). (b) High-resolution image showing the Ge/Si and GeSn/Ge interfaces, marked by white dashed lines. (c) Image showing the relaxation start point in GeSn layer, marked by red dashed line.

Fig. 4 shows the TEM images of sample S2 taken at different resolution levels. Clear interfaces between Ge/Si and GeSn/Ge can be observed in the images. Additionally, two white lines are also observed near the GeSn/Ge interface, which are caused by contrast variations, induced by the strain fields at the interface. The white contrast lines in DF-images observed throughout the entire epitaxial layer are attributed to dislocations. As observed in Fig. 4(a-c), a large number of misfit dislocations form at the Ge/Si interface, and some of these dislocations propagate upward, generating threading dislocations. In the higher-resolution image in Fig. 4(b), it can be observed that a large number of dislocations are present in the region near the Ge/Si interface, while fewer dislocations are found in the upper Ge buffer layer. Which means most of the dislocations are confined to the low-temperature Ge layer. However, a significant number of dislocations still exist in the upper Ge region. This is due to the insufficient thickness of the Ge buffer and the low annealing temperature. Additionally, it can be observed that the threading dislocation density in the GeSn layer near the interface is comparable to the upper Ge buffer layer. The threading dislocations propagate through the interface and extend into the GeSn layer, as shown in Fig. 4(b). However, as shown in Fig. 4(c), additional threading dislocations are generated approximately 60 nm (red dashed line) from the interface, indicating stress relaxation at this location. This is consistent with the two GeSn peaks observed in the XRD measurements. After this, the dislocations propagate upward, undergoing merging and annihilation. Similar phenomenon has also been reported in the work of other research groups[16]. Unlike the relaxation mechanism in CVD-grown GeSn on Si samples, where 60° dislocation half-loops are generated after the critical thickness and propagate downward to release strain, this process subsequently enables dislocation-free GeSn growth in the upper layer[25]. The different relaxation mechanism results in a higher threading dislocation density in MBE-grown GeSn, which consequently affects its optical properties.

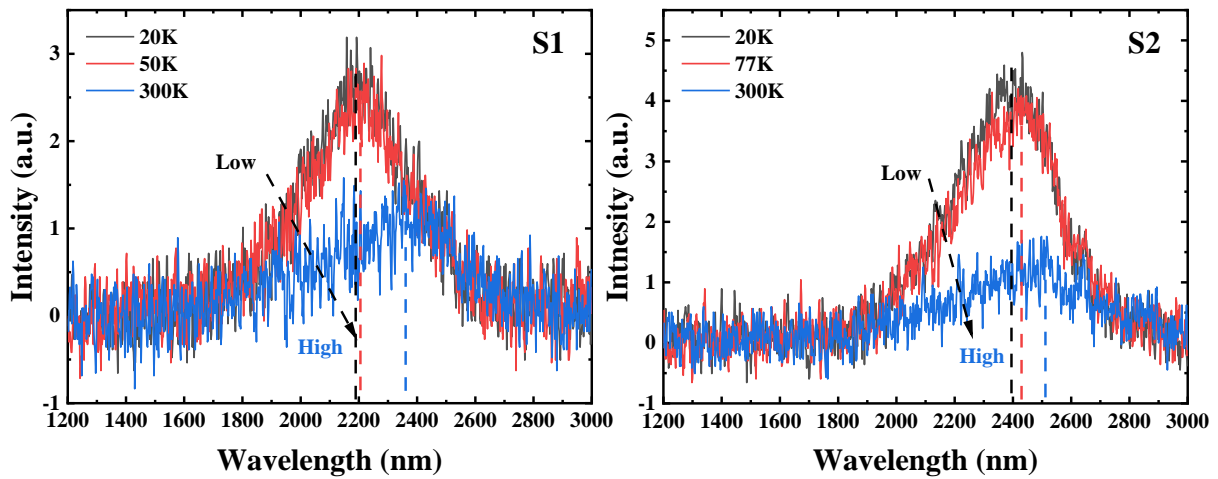

Figure 5. Temperature dependent PL of the GeSn samples with Sn content of (a) 9.2% and (b) 11.4%.

PL spectroscopy serves as a fundamental technique for assessing the optical properties of GeSn alloys, providing crucial insights into their band structure and carrier recombination dynamics. The transition from an indirect to a direct bandgap in GeSn occurs at a critical Sn concentration, typically within the range of 6–10%, depending on strain conditions. Fig. 5 shows the temperature-dependent PL measurements conducted on GeSn films with 9.2% and 11.4% Sn content. The presence of direct bandgap emission is evidenced by the observation of a single, dominant PL peak at the expected wavelength given the Sn content of each sample[26].

The results demonstrate strong mid-infrared emission, in agreement with previous reports of room-temperature PL extending up to 2230 nm for GeSn alloys with approximately 10% Sn[5]. The observed narrowing of the PL peak with decreasing temperature suggests a reduction in non-radiative recombination processes, further corroborating the direct bandgap nature of the material[27,28]. The high radiative efficiency observed in these films indicates effective strain relaxation, which mitigates defect-induced non-radiative pathways and enhances optical performance. Also, two samples at different Sn content were used to demonstrate tuning the PL emission wavelength as a function of Sn content.

Comparison with prior studies reveals consistency in the indirect-to-direct bandgap transition mechanism. GeSn alloys with lower Sn concentrations typically exhibit broadened PL spectra due to the coexistence of direct and indirect bandgap transitions, whereas GeSn samples with Sn compositions exceeding 10% display a sharp, well-defined peak indicative of a fully direct bandgap. These findings underscore the significance of strain engineering and defect minimization in optimizing the optical properties of GeSn.

**Conclusion:**

In summary, this study demonstrates the first observation of direct bandgap photoluminescence from GeSn films grown on Si(100) substrates using MBE without the need for post-growth annealing. High-quality GeSn layers with Sn contents of 9.2% and 11.4% were successfully synthesized, showing efficient strain relaxation and uniform Sn incorporation. Structural characterizations via XRD, SIMS, AFM, and TEM confirmed the high-quality of the films and the absence of significant Sn segregation. Temperature-dependent PL measurements unequivocally confirmed the direct bandgap nature of the GeSn films, exhibiting a well-defined emission peak with high radiative efficiency. These findings establish MBE as a viable and CMOS-compatible approach for fabricating high-Sn-content GeSn films with direct bandgap characteristics, paving the way for their integration into silicon-based optoelectronic devices such as lasers and detectors.

**Acknowledgement:**

The authors thank Dr. Mohammad Zamani Alavijeh for AFM measurement. This work was supported by the Multidisciplinary University Research Initiative (MURI) Program through U.S. Air Force Office of Scientific Research (AFOSR) Grant No. FA9550-19-1-0341.

**Conflict of Interest:**

There is no conflict of interest.

**References:**

1. Wirths, S.; Buca, D.; Mantl, S., Si–Ge–Sn alloys: From growth to applications. *Progress in Crystal Growth and Characterization of Materials* **2016,** *62* (1), 1-39.
2. Zhou, Y.; Dou, W.; Du, W.; Pham, T.; Ghetmiri, S. A.; Al-Kabi, S.; Mosleh, A.; Alher, M.; Margetis, J.; Tolle, J.; Sun, G.; Soref, R.; Li, B.; Mortazavi, M.; Naseem, H.; Yu, S.-Q., Systematic study of GeSn heterostructure-based light-emitting diodes towards mid-infrared applications. *Journal of Applied Physics* **2016,** *120* (2), 023102.
3. von den Driesch, N.; Stange, D.; Wirths, S.; Mussler, G.; Holländer, B.; Ikonic, Z.; Hartmann, J. M.; Stoica, T.; Mantl, S.; Grützmacher, D.; Buca, D., Direct Bandgap Group IV Epitaxy on Si for Laser Applications. *Chemistry of Materials* **2015,** *27* (13), 4693-4702.
4. Polak, M. P.; Scharoch, P.; Kudrawiec, R., The electronic band structure of Ge1−xSnx in the full composition range: indirect, direct, and inverted gaps regimes, band offsets, and the Burstein–Moss effect. *Journal of Physics D: Applied Physics* **2017,** *50* (19), 195103.
5. Ghetmiri, S. A.; Du, W.; Margetis, J.; Mosleh, A.; Cousar, L.; Conley, B. R.; Domulevicz, L.; Nazzal, A.; Sun, G.; Soref, R. A.; Tolle, J.; Li, B.; Naseem, H. A.; Yu, S.-Q., Direct-bandgap GeSn grown on silicon with 2230 nm photoluminescence. *Applied Physics Letters* **2014,** *105* (15), 151109.
6. Zheng, J.; Liu, Z.; Xue, C.; Li, C.; Zuo, Y.; Cheng, B.; Wang, Q., Recent progress in GeSn growth and GeSn-based photonic devices. *Journal of Semiconductors* **2018,** *39* (6).
7. Miao, Y.; Wang, G.; Kong, Z.; Xu, B.; Zhao, X.; Luo, X.; Lin, H.; Dong, Y.; Lu, B.; Dong, L.; Zhou, J.; Liu, J.; Radamson, H. H., Review of Si-Based GeSn CVD Growth and Optoelectronic Applications. *Nanomaterials (Basel)* **2021,** *11* (10).
8. Wang, W.; Zhou, Q.; Dong, Y.; Tok, E. S.; Yeo, Y.-C., Critical thickness for strain relaxation of Ge1−xSnx (x ⩽ 0.17) grown by molecular beam epitaxy on Ge(001). *Applied Physics Letters* **2015,** *106* (23), 232106.
9. Wang, N.; Xue, C.; Wan, F.; Zhao, Y.; Xu, G.; Liu, Z.; Zheng, J.; Zuo, Y.; Cheng, B.; Wang, Q., Spontaneously Conversion from Film to High Crystalline Quality Stripe during Molecular Beam Epitaxy for High Sn Content GeSn. *Scientific Reports* **2020,** *10* (1), 6161.
10. Rathore, J.; Nanwani, A.; Mukherjee, S.; Das, S.; Moutanabbir, O.; Mahapatra, S., Composition uniformity and large degree of strain relaxation in MBE-grown thick GeSn epitaxial layers, containing 16% Sn. *Journal of Physics D: Applied Physics* **2021,** *54* (18), 185105.
11. Pukite, P. R.; Harwit, A.; Iyer, S. S., Molecular beam epitaxy of metastable, diamond structure SnxGe1−x alloys. *Applied Physics Letters* **1989,** *54* (21), 2142-2144.
12. M. Eldose, N.; Stanchu, H.; Das, S.; Bikmukhametov, I.; Li, C.; Shetty, S.; Mazur, Y. I.; Yu, S.-Q.; Salamo, G. J., Strain-Mediated Sn Incorporation and Segregation in Compositionally Graded Ge1–xSnx Epilayers Grown by MBE at Different Temperatures. *Crystal Growth & Design* **2023,** *23* (11), 7737-7743.
13. Kormoš, L.; Kratzer, M.; Kostecki, K.; Oehme, M.; Šikola, T.; Kasper, E.; Schulze, J.; Teichert, C., Surface analysis of epitaxially grown GeSn alloys with Sn contents between 15% and 18%. *Surface and Interface Analysis* **2017,** *49* (4), 297-302.
14. Oehme, M.; Kostecki, K.; Schmid, M.; Oliveira, F.; Kasper, E.; Schulze, J., Epitaxial growth of strained and unstrained GeSn alloys up to 25% Sn. *Thin Solid Films* **2014,** *557*, 169-172.
15. Qian, K.; Wu, S.; Qian, J.; Yang, K.; An, Y.; Cai, H.; Lin, G.; Wang, J.; Xu, J.; Huang, W.; Chen, S.; Li, C., Secondary epitaxy of high Sn fraction GeSn layer on strain-relaxed GeSn virtue substrate by molecular beam epitaxy. *Journal of Physics D: Applied Physics* **2023,** *56* (7).
16. Wan, F.; Xu, C.; Wang, X.; Xu, G.; Cheng, B.; Xue, C., Study of strain evolution mechanism in Ge1−xSnx materials grown by low temperature molecular beam epitaxy. *Journal of Crystal Growth* **2022,** *577*.
17. Kondratenko, S. V.; Derenko, S. S.; Mazur, Y. I.; Stanchu, H.; Kuchuk, A. V.; Lysenko, V. S.; Lytvyn, P. M.; Yu, S. Q.; Salamo, G. J., Impact of defects on photoexcited carrier relaxation dynamics in GeSn thin films. *Journal of Physics: Condensed Matter* **2020,** *33* (6), 065702.

18. Giunto, A.; Fontcuberta i Morral, A., Defects in Ge and GeSn and their impact on optoelectronic properties. *Applied Physics Reviews* **2024,** *11* (4), 041333.
19. Shevchenko, S.; Tereshchenko, A., Peculiarities of dislocation photoluminescence in germanium with quasi-equilibrium dislocation structure. *physica status solidi c* **2007,** *4* (8), 2898-2902.
20. Amoah, S.; Ojo, S.; Tran, H.; Abernathy, G.; Zhou, Y.; Du, W.; Margetis, J.; Tolle, J.; Li, B.; Yu, S.-Q. In *Electrically Injected GeSn Laser on Si Operating up to 110K*, Conference on Lasers and Electro-Optics, San Jose, California, 2021/05/09; Kang, J. T. S. I. I. M. D. L. N. P. S. P. V. N. J. D. C. F. T. G. Q.; Saraceno, C., Eds. Optica Publishing Group: San Jose, California, 2021; p SM1H.4.
21. Rosson, N.; Acharya, S.; Fischer, A. M.; Collier, B.; Ali, A.; Torabi, A.; Du, W.; Yu, S.-Q.; Scott, R. C., Development of GeSn epitaxial films with strong direct bandgap luminescence in the mid-wave infrared region using a commercial chemical vapor deposition reactor. *Journal of Vacuum Science & Technology B* **2024,** *42* (5), 052210.
22. Acharya, S.; Stanchu, H.; Kumar, R.; Ojo, S.; Alher, M.; Benamara, M.; Chang, G. E.; Li, B.; Du, W.; Yu, S. Q., Electrically Injected Mid-Infrared GeSn Laser on Si Operating at 140 K. *IEEE Journal of Selected Topics in Quantum Electronics* **2025,** *31* (1: SiGeSn Infrared Photon. and Quantum Electronics), 1-7.
23. Lin, G.; Qian, K.; Cai, H.; Zhao, H.; Xu, J.; Chen, S.; Li, C.; Hickey, R.; Kolodzey, J.; Zeng, Y., Enhanced photoluminescence of GeSn by strain relaxation and spontaneous carrier confinement through rapid thermal annealing. *Journal of Alloys and Compounds* **2022,** *915*.
24. Zhang, D.; Lu, J.; Liu, Z.; Wan, F.; Liu, X.; Pang, Y.; Zhu, Y.; Cheng, B.; Zheng, J.; Zuo, Y.; Xue, C., Sharp interface of undoped Ge/SiGe quantum well grown by ultrahigh vacuum chemical vapor deposition. *Applied Physics Letters* **2022,** *121* (2).
25. Dou, W.; Benamara, M.; Mosleh, A.; Margetis, J.; Grant, P.; Zhou, Y.; Al-Kabi, S.; Du, W.; Tolle, J.; Li, B.; Mortazavi, M.; Yu, S.-Q., Investigation of GeSn Strain Relaxation and Spontaneous Composition Gradient for Low-Defect and High-Sn Alloy Growth. *Scientific Reports* **2018,** *8* (1), 5640.
26. Song, Z.; Fan, W.; Tan, C. S.; Wang, Q.; Nam, D.; Zhang, D. H.; Sun, G., Band structure of $Ge_{1-x}Sn_x$ alloy: a full-zone 30-band k · p model. *New Journal of Physics* **2019,** *21* (7), 073037.
27. Sasaki, A.; Nishizuka, K.; Wang, T.; Sakai, S.; Kaneta, A.; Kawakami, Y.; Fujita, S., Radiative carrier recombination dependent on temperature and well width of InGaN/GaN single quantum well. *Solid State Communications* **2004,** *129* (1), 31-35.
28. Lamba, S.; Joshi, S. K., Temperature dependent photoluminescence in self assembled InAs quantum dot arrays. *physica status solidi (b)* **2003,** *239* (2), 353-360.